\documentclass[
reprint,
superscriptaddress,
amsmath,amssymb,
aps,
prb,
]{revtex4-2}

\usepackage{xcolor}
\usepackage{multirow}
\usepackage{longtable}
\usepackage{graphicx}
\usepackage{dcolumn}
\usepackage{bm}
\usepackage{hyperref}
\newcolumntype{?}{!{\vrule width 1pt}}
\usepackage{float}

\usepackage{booktabs}
\usepackage{amsmath}
\setcitestyle{super}

\begin{document}

\title{Synthesizability and Mechanical Properties of High-Entropy Borides:\\ First-Principles and Machine Learning Studies}

\author{Luke Moore}
\thanks{The two authors contributed equally to this work.}
\affiliation{Department of Physics, University of Alabama at Birmingham, Birmingham, Alabama 35294, USA}

\author{Ethan Fox}
\thanks{The two authors contributed equally to this work.}
\affiliation{Department of Physics, University of Alabama at Birmingham, Birmingham, Alabama 35294, USA}

\author{Bria Storr}
\affiliation{Otto H. York Department of Chemical and Materials Engineering, New Jersey Institute of Technology, Newark, New Jersey 07102, USA}

\author{Jayden R. Palomino}
\affiliation{Department of Physics and Astronomy, Mississippi State University, Starkville, Mississippi 39762, USA}

\author{Shane A. Catledge}
\affiliation{Department of Physics, University of Alabama at Birmingham, Birmingham, Alabama 35294, USA}

\author{Yogesh K. Vohra}
\affiliation{Department of Physics, University of Alabama at Birmingham, Birmingham, Alabama 35294, USA}

\author{Cheng-Chien Chen}
\email{Corresponding author: chencc@uab.edu}
\affiliation{Department of Physics, University of Alabama at Birmingham, Birmingham, Alabama 35294, USA}

\date{\today}

\begin{abstract}
We perform density functional theory (DFT) calculations to investigate five-metal high-entropy borides (HEBs) in the hexagonal AlB$_2$ structure, considering all 126 possible elemental combinations among the nine group 4-6 transition metals (Ti, V, Cr, Zr, Nb, Mo, Hf, Ta, and W). The entropy forming ability (EFA) descriptor is employed to evaluate their single-phase synthesizability, and the resulting EFA predictions show good agreement with the experimental data for selected HEBs. Mechanical properties are computed using special quasi-random structures. Several mechanically unstable compounds -- primarily those containing Cr -- are also predicted to be less synthesizable. Machine learning (ML) models are developed to analyze the results. This combined {\it ab initio} and ML study provides a systematic roadmap for identifying mechanically superior single-phase HEBs.
\end{abstract}


\maketitle

\section{Introduction}
Transition-metal (TM) borides are an important class of ultra-high-temperature ceramics known for their exceptional hardness, incompressibility, and chemical stability, making them attractive for applications under extreme conditions~\cite{zhang2018inherent,golla2020review,wyatt2024ultra}. Their superior mechanical properties originate from strong covalent boron-boron bonds together with robust metal-boron bonding. More recently, high-entropy borides (HEBs) have emerged as a promising subclass of compositionally complex ceramics~\cite{gild2016high,oses2020high,wright2020high,feng2021high,dube2024underpinning,qureshi2024high}. High-entropy materials consist of five or more elements in near-equimolar ratios, and they can be stabilized by configurational entropy and exhibit properties beyond those of their binary or ternary counterparts~\cite{yeh2004nanostructured,cantor2004microstructural,hsu2024clarifying}. Recent experiments have successfully synthesized a variety of HEBs and demonstrated their excellent hardness, thermal stability, and oxidation resistance~\cite{gild2016high,feng2021superhard,iwan2022high,storr2022properties,storr2023high,yang2023effect}. These advances suggest that HEBs represent a rich materials-design space for next-generation extreme applications.

Despite their promise, not all elemental combinations can form stable single-phase HEBs. The vast compositional space makes experimental trial-and-error approaches prohibitively expensive for systematically identifying synthesizable HEBs. Consequently, reliable computational descriptors are highly desirable. Several metrics, including entropy forming ability (EFA)~\cite{sarker2018high}, disordered enthalpy-entropy descriptor (DEED)~\cite{divilov2024disordered}, mixed enthalpy-entropy descriptor (MEED)~\cite{dey2024mixed}, as well as designing principles based on lattice mismatch and valence electron count~\cite{kretschmer2024explaining,gu2025robust} have shown success in predicting the synthesizability of high-entropy ceramics. Comprehensive assessments across large families of compositions can thereby accelerate material discovery.

Beyond synthesizability, predicting the mechanical properties of HEBs and understanding the effects of disorder remain challenging. Accurate first-principles modeling typically requires large supercells to capture their disordered atomic arrangements. The special quasi-random structure (SQS) approach~\cite{zunger1990special,van2002alloy} has emerged as an effective framework for representing substitutional disorder in periodic density functional theory (DFT) calculations and has successfully reproduced the structural and mechanical properties of various high-entropy ceramics~\cite{wang2018ab,yang2018structural,iwan2022high,xiong2022frist,xiao2022ab,storr2023high,mo2026chromium}. However, comprehensive datasets for HEBs remain limited, motivating further systematic investigations and data-driven materials design efforts. Such SQS-based studies can provide valuable insights into the composition-structure-property relationships of HEBs.
 
In this work, we combine EFA analysis, DFT-SQS calculations, and machine-learning (ML) techniques to systematically study all 126 five-metal HEBs formed from the nine group 4-6 transition metals (Ti, V, Cr, Zr, Nb, Mo, Hf, Ta, and W). We compute their synthesizability and mechanical properties, and compare the results with available experimental data. The calculations reveal that several Cr-containing compounds exhibit large discrepancies in the elastic moduli between the Voigt and Reuss approximations~\cite{hill1952elastic,ravindran1998density} -- indicative of strong elastic anisotropy or mechanical instability -- and they also tend to exhibit low EFA values and reduced synthesizability. We further employ ML models to analyze the DFT dataset and identify key compositional and structural descriptors governing the HEB properties. In particular, lattice-mismatch-related features associated with bond-length distributions are found to be correlated with mechanical strength, highlighting the role of disorder. Together, these results provide a systematic roadmap for identifying synthesizable HEBs with superior mechanical properties for applications in extreme environments.


\section{Computational Methods}\label{methods}

\subsection{First-Principles Calculation}

First-principles calculations are performed using the Vienna Ab Initio Simulation Package (VASP)~\cite{kresse1993ab,kresse1996efficient}, a plane-wave pseudopotential implementation of density functional theory (DFT). The Perdew-Burke-Ernzerhof generalized gradient approximation (PBE-GGA) exchange-correlation functional is employed~\cite{perdew1996generalized}. The electronic self-consistent field calculations are converged to within $10^{-6}$ eV, while ionic relaxations are performed until the residual forces on all atoms are smaller than $10^{-2}$ eV/\AA. A plane-wave energy cutoff of 450 eV is used, corresponding to at least 1.4 times the maximum recommended cutoff among the constituent pseudopotentials. Brillouin-zone integrations are performed using $\Gamma$-centered $k$-point meshes with a k-points-per-reciprocal-atom (KPPRA) value of 6000, automatically generated by the AFLOW package~\cite{curtarolo2012aflow} for each input structure.

To evaluate the synthesizability of hexagonal HEBs in the AlB$_2$ structure, DFT calculations are carried out for the minimal 15-atom unit cells generated by the AFLOW-Partial OCCupation (AFLOW-POCC) algorithm~\cite{yang2016modeling,oses2023aflow++}. Each unit cell contains five transition-metal atoms and ten boron atoms. AFLOW-POCC generates a set of distinct 15-atom structures that represent the specified stoichiometry and crystallographic symmetries. The synthesizability can then be assessed from the resulting energy distribution of these distinct POCC structures through the EFA descriptor discussed later.
 
To evaluate the mechanical properties, DFT calculations are performed using the special quasi-random structure (SQS)~\cite{zunger1990special,van2002alloy}, which provides the best periodic-supercell approximation to a fully disordered state for a given number of atoms. For each five-metal HEB composition, a 60-atom SQS supercell is first fully relaxed, followed by calculations of the elastic constants using the VASP stress-strain method. The resulting elastic constants are then used to evaluate the mechanical properties based on the Voigt-Reuss-Hill approximation~\cite{hill1952elastic,ravindran1998density}. Previous studies on HEBs have shown that larger SQS supercells up to 150 atoms do not lead to substantial differences in the calculated mechanical properties compared to those obtained using 60-atom supercells~\cite{storr2023high}.

\subsection{Machine Learning Simulation}
The mechanical properties computed by DFT are further utilized to develop machine-learning (ML) regression models. For each HEB composition, such as CrHfMoNbTaB$_{10}$, a set of 60 composition descriptors is generated~\cite{ward2016general,chen2021machine}. These descriptors comprise four categories: stoichiometric, elemental-property-based, valence-orbital, and ionic-bonding attributes. The stoichiometric attributes (3 total) quantify compositional diversity through the $L_p$ norms of the elemental fractions ($p=0,2,3$). The elemental-property-based attributes (50 total) represent statistical descriptors (minimum, maximum, range, fractional weighted mean, and deviation) of ten elemental properties: atomic number, atomic mass, periodic-table row and column, atomic radius, electronegativity, and the numbers of valence $s$, $p$, $d$, and $f$ electrons. The valence-orbital attributes (4 total) quantify the fractional occupations of the $s$, $p$, $d$, and $f$ orbitals. The ionic-bonding attributes (3 total) characterize the potential ionic character through charge-balance and electronegativity-based metrics. These compositional features are obtained using the Python Materials Genomics (pymatgen) package~\cite{ong2013python}. 

Additional 19 structural features based on the fully relaxed 60-atom SQS supercells are further incorporated to improve the ML models. These structural features include nearest-neighbor (NN) distances extracted from the supercells and are grouped into six categories: metal-boron (M-B), metal-metal along $c$-axis (M-M$\perp$), boron-boron along $c$-axis (B-B$\perp$), boron-boron within $ab$-plane (B-B$\parallel$), metal-metal within $ab$-plane (M-M$\parallel$), and the second NN boron-boron within $ab$-plane plane (B-B 2nd$\parallel$). For each category, the mean ($\mu$), standard deviation ($\sigma$), and range of distances are computed; the relaxed cell volume is also included, yielding a total of 19 DFT derived structural descriptors.

To perform the regression task, we employ the Extreme Gradient Boosting (XGBoost) algorithm implemented in the XGBoost library~\cite{chen2016xgboost}. Gradient-boosted tree methods construct an ensemble of decision trees sequentially, with each new tree trained to reduce the prediction errors of the existing ensemble by minimizing a specified loss function. XGBoost extends traditional gradient boosting by incorporating both $L_1$ and $L_2$ regularization terms and utilizing second-order gradient information during optimization. These features improve model generalization, reduce overfitting, and enhance computational efficiency. XGBoost has been widely applied in materials informatics~\cite{choi2019data,smith2023machine,bakhtiari2024xgboost,pang2026physics} and is well suited for small datasets considered in this study. Furthermore, XGBoost provides feature-importance metrics that facilitate physical interpretation of the ML models, allowing us to identify descriptors strongly correlated with the target properties.

For model training, the dataset is divided into an 80\% train-validation set and a 20\% test set. The train-validation set is used to determine the optimal hyperparameters through a grid-search procedure combined with four-fold cross-validation. To mitigate overfitting on the relatively small DFT dataset, we explore a range of hyperparameter settings, including the number of trees, learning rate, and maximum tree depth. Additional hyperparameters related to data subsampling and node splitting are also optimized. Model performance is evaluated using the Pearson correlation coefficient ($r$) and the mean absolute error (MAE), as discussed below.

\begin{figure*}[tbp]
\centering
\includegraphics[width=\textwidth]{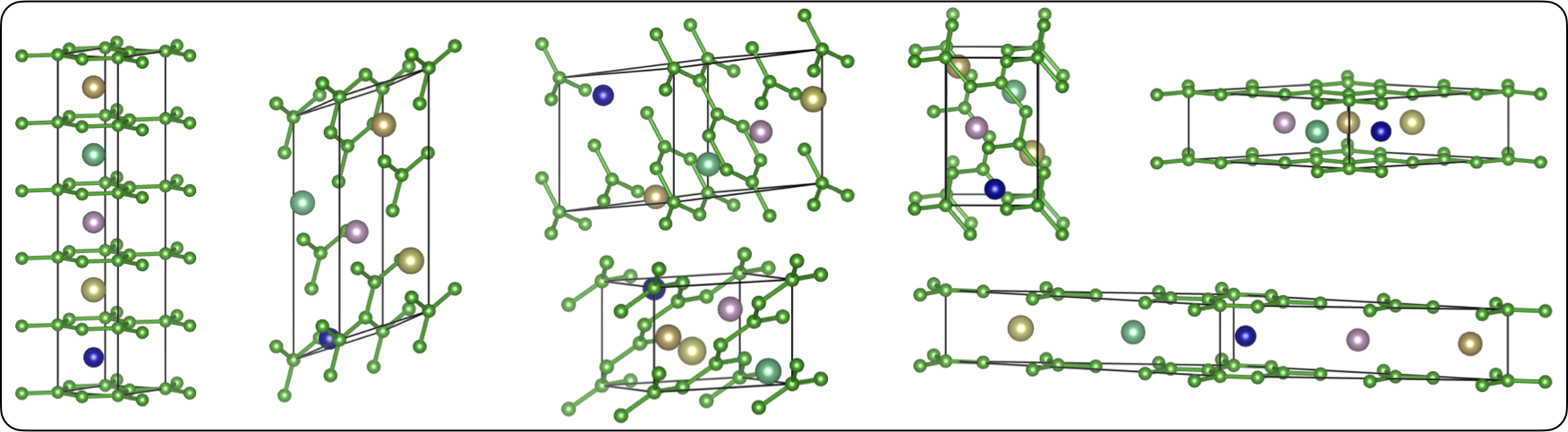}
\caption{Symmetry-distinct 15-atom unit cells for five-metal high-entropy borides in the hexagonal AlB$_2$ phase, generated using the AFLOW Partial OCCupation (AFLOW-POCC) algorithm~\cite{yang2016modeling,oses2023aflow++}. Green spheres represent boron atoms, while larger spheres denote transition-metal atoms. The structures are visualized using the VESTA software~\cite{momma2011vesta}.
}
\label{fig:pocc}
\end{figure*}

\section{Results and Discussion}\label{results}

\subsection{Entropy Forming Ability and Synthesizability}

For a hexagonal five-metal HEB, the minimal 15-atom unit cell yields 84 POCC structures generated by the AFLOW-POCC algorithm~\cite{yang2016modeling,oses2023aflow++}. Representative symmetry-distinct POCC configurations are shown in Fig. \ref{fig:pocc}. The total energy of each structure is computed using DFT, and the resulting energy distribution is used to compute the entropy-forming ability (EFA)~\cite{sarker2018high}. EFA is evaluated as the inverse of the energy dispersion $\sigma_E$:
\begin{eqnarray}
\textrm{EFA} &\equiv& \sigma^{-1}_E,\nonumber\\
\sigma_E&\equiv&
\sqrt{
\frac{
\sum_i g_i\left(E_i-\langle E\rangle\right)^2
}{
\left(\sum_i g_i\right)-1
}
}.
\end{eqnarray}
Here, $E_i$ is the DFT total energy of the $i$-th POCC structure and $g_i$ is its degeneracy (which is equal to 10 for the 15-atom hexagonal five-metal HEB phase). $\langle E \rangle \equiv \sum_i g_i E_i /\sum_i g_i$ is the average energy. A large EFA corresponds to a narrow energy distribution, indicating that many atomic arrangements are energetically comparable and therefore more likely to form a single phase. Conversely, a small EFA reflects a broad energy distribution and a stronger tendency towards phase separation.

\begin{figure*}[tbp]
\centering
\includegraphics[width=\textwidth]{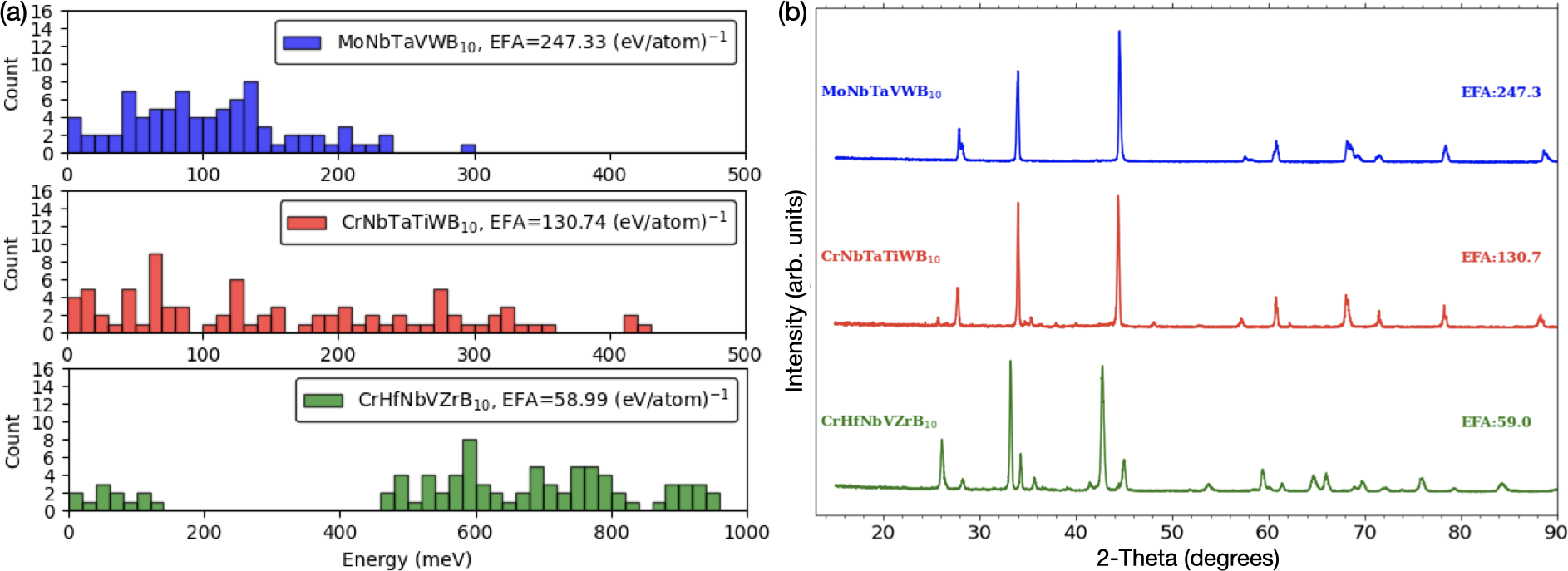}
\caption{(a) Histograms of the total energies of POCC structures for three representative high-entropy borides with high (MoNbTaVWB$_{10}$), intermediate (CrNbTaTiWB$_{10}$), and low (CrHfNbVZrB$_{10}$) EFA values. Energy zero corresponds to the lowest-energy structure of each composition. Higher EFA values are associated with narrower energy distributions. (b) Experimental x-ray diffraction patterns of the corresponding compounds. High-EFA compounds exhibit fewer and sharper diffraction peaks, indicating a greater propensity for single-phase synthesis. The calculated EFA values agree well with the experiments.
}
\label{fig:efa}
\end{figure*}

Figure \ref{fig:efa}(a) shows energy histograms of the POCC structures for three representative HEBs: MoNbTaVWB$_{10}$, CrNbTaTiWB$_{10}$, and CrHfNbVZrB$_{10}$. The energy range of CrHfNbVZrB$_{10}$ is approximately 1000 meV, which is more than twice that of the other two compounds. This results in a substantially broader energy distribution $\sigma_E$ of CrHfNbVZrB$_{10}$ and a lower EFA. The computed EFA values are 247.33, 130.74, and 58.99 (eV/atom)$^{-1}$ for MoNbTaVWB$_{10}$, CrNbTaTiWB$_{10}$, and CrHfNbVZrB$_{10}$, respectively.

To evaluate the predictive capability of the EFA descriptor for single-phase formation in high-entropy materials, we compare the EFA values with the experimental observations obtained in this study. Figure \ref{fig:efa}(b) shows the x-ray diffraction (XRD) patterns of MoNbTaVWB$_{10}$, CrNbTaTiWB$_{10}$, and CrHfNbVZrB$_{10}$, which exhibit high, intermediate, and low EFA values, respectively. Among the three HEBs, the XRD data follow the predicted trend. MoNbTaVWB$_{10}$, which has the highest EFA, exhibits fewer and sharper diffraction peaks, consistent with a greater propensity for single-phase formation. CrNbTaTiWB$_{10}$, with an intermediate EFA value, shows several additional weak peaks. Finally, CrHfNbVZrB$_{10}$, which has the lowest EFA, exhibits multiple extra peaks indicative of secondary phases. These observations are consistent with the predictions and demonstrates that EFA is a useful metric for single-phase synthesizability.

Table \ref{table:efa} summarizes the calculated EFA values for all 126 five-metal HEBs studied. The EFA values span a wide range, from 58.99 to 294.64 (eV/atom)$^{-1}$, indicating substantial variations among different compositions. The five compounds with the highest EFA values are HfMoNbTaZrB$_{10}$ (294.64), MoNbTaVWB$_{10}$ (247.33), HfNbTaTiZrB$_{10}$ (240.56), HfMoNbTaTiB$_{10}$ (231.49), and MoNbTaTiVB$_{10}$ (231.02). Notably, the highest-EFA compound HfMoNbTaZrB$_{10}$ has been synthesized as a single-phase HEB in earlier work~\cite{iwan2022high,storr2022properties}, providing additional experimental support for the predictive capability of the EFA descriptor. In contrast, the five lowest-EFA compounds are CrHfNbVZrB$_{10}$ (58.99), CrHfTaVZrB$_{10}$ (62.83), CrHfTiVZrB$_{10}$ (62.98), CrNbTaVZrB$_{10}$ (73.95), and CrHfNbTiZrB$_{10}$ (75.90). A notable trend is that all five lowest-EFA compounds contain Cr, whereas none of the five highest-EFA compounds contain Cr. This observation suggests that the presence of Cr is often associated with reduced EFA values and a lower propensity for single-phase HEB formation.

\begin{table*}[t]
\caption{Entropy-forming ability (EFA) values in (eV/atom)$^{-1}$ for the 126 five-metal high-entropy borides studied.}
\label{table:efa}
\centering
\small
\setlength{\arrayrulewidth}{0.8pt}
\resizebox{\textwidth}{!}{%
\begin{tabular}{lr|lr|lr|lr}
\toprule
Material & EFA & Material & EFA & Material & EFA & Material & EFA\\
\midrule
CrHfMoNbTaB$_{10}$ & 107.75 & CrHfTiVZrB$_{10}$ & 62.98 & CrNbVWZrB$_{10}$ & 107.27 & HfNbTiVWB$_{10}$ & 155.65 \\
CrHfMoNbTiB$_{10}$ & 137.39 & CrHfTiWZrB$_{10}$ & 122.22 & CrTaTiVWB$_{10}$ & 131.08 & HfNbTiVZrB$_{10}$ & 116.69 \\
CrHfMoNbVB$_{10}$ & 108.24 & CrHfVWZrB$_{10}$ & 87.23 & CrTaTiVZrB$_{10}$ & 88.64 & HfNbTiWZrB$_{10}$ & 164.45 \\
CrHfMoNbWB$_{10}$ & 125.06 & CrMoNbTaTiB$_{10}$ & 153.09 & CrTaTiWZrB$_{10}$ & 106.39 & HfNbVWZrB$_{10}$ & 142.30 \\
CrHfMoNbZrB$_{10}$ & 90.75 & CrMoNbTaVB$_{10}$ & 166.19 & CrTaVWZrB$_{10}$ & 100.94 & HfTaTiVWB$_{10}$ & 135.09 \\
CrHfMoTaTiB$_{10}$ & 129.82 & CrMoNbTaWB$_{10}$ & 160.73 & CrTiVWZrB$_{10}$ & 114.60 & HfTaTiVZrB$_{10}$ & 129.22 \\
CrHfMoTaVB$_{10}$ & 106.81 & CrMoNbTaZrB$_{10}$ & 98.85 & HfMoNbTaTiB$_{10}$ & 231.49 & HfTaTiWZrB$_{10}$ & 143.09 \\
CrHfMoTaWB$_{10}$ & 116.31 & CrMoNbTiVB$_{10}$ & 193.48 & HfMoNbTaVB$_{10}$ & 181.08 & HfTaVWZrB$_{10}$ & 133.44 \\
CrHfMoTaZrB$_{10}$ & 91.25 & CrMoNbTiWB$_{10}$ & 124.57 & HfMoNbTaWB$_{10}$ & 185.13 & HfTiVWZrB$_{10}$ & 142.02 \\
CrHfMoTiVB$_{10}$ & 121.19 & CrMoNbTiZrB$_{10}$ & 121.33 & HfMoNbTaZrB$_{10}$ & 294.64 & MoNbTaTiVB$_{10}$ & 231.02 \\
CrHfMoTiWB$_{10}$ & 98.06 & CrMoNbVWB$_{10}$ & 170.07 & HfMoNbTiVB$_{10}$ & 171.49 & MoNbTaTiWB$_{10}$ & 157.72 \\
CrHfMoTiZrB$_{10}$ & 101.38 & CrMoNbVZrB$_{10}$ & 94.92 & HfMoNbTiWB$_{10}$ & 132.90 & MoNbTaTiZrB$_{10}$ & 218.48 \\
CrHfMoVWB$_{10}$ & 109.55 & CrMoNbWZrB$_{10}$ & 116.90 & HfMoNbTiZrB$_{10}$ & 222.30 & MoNbTaVWB$_{10}$ & 247.33 \\
CrHfMoVZrB$_{10}$ & 76.61 & CrMoTaTiVB$_{10}$ & 171.52 & HfMoNbVWB$_{10}$ & 169.11 & MoNbTaVZrB$_{10}$ & 153.82 \\
CrHfMoWZrB$_{10}$ & 90.68 & CrMoTaTiWB$_{10}$ & 115.52 & HfMoNbVZrB$_{10}$ & 132.93 & MoNbTaWZrB$_{10}$ & 186.69 \\
CrHfNbTaTiB$_{10}$ & 103.34 & CrMoTaTiZrB$_{10}$ & 115.63 & HfMoNbWZrB$_{10}$ & 143.72 & MoNbTiVWB$_{10}$ & 155.75 \\
CrHfNbTaVB$_{10}$ & 81.32 & CrMoTaVWB$_{10}$ & 158.82 & HfMoTaTiVB$_{10}$ & 165.62 & MoNbTiVZrB$_{10}$ & 144.23 \\
CrHfNbTaWB$_{10}$ & 123.87 & CrMoTaVZrB$_{10}$ & 94.66 & HfMoTaTiWB$_{10}$ & 124.38 & MoNbTiWZrB$_{10}$ & 131.86 \\
CrHfNbTaZrB$_{10}$ & 80.88 & CrMoTaWZrB$_{10}$ & 108.68 & HfMoTaTiZrB$_{10}$ & 207.65 & MoNbVWZrB$_{10}$ & 154.10 \\
CrHfNbTiVB$_{10}$ & 96.23 & CrMoTiVWB$_{10}$ & 128.37 & HfMoTaVWB$_{10}$ & 155.73 & MoTaTiVWB$_{10}$ & 142.41 \\
CrHfNbTiWB$_{10}$ & 122.63 & CrMoTiVZrB$_{10}$ & 101.63 & HfMoTaVZrB$_{10}$ & 136.83 & MoTaTiVZrB$_{10}$ & 143.38 \\
CrHfNbTiZrB$_{10}$ & 75.90 & CrMoTiWZrB$_{10}$ & 95.98 & HfMoTaWZrB$_{10}$ & 136.54 & MoTaTiWZrB$_{10}$ & 123.00 \\
CrHfNbVWB$_{10}$ & 115.72 & CrMoVWZrB$_{10}$ & 98.91 & HfMoTiVWB$_{10}$ & 116.68 & MoTaVWZrB$_{10}$ & 144.37 \\
CrHfNbVZrB$_{10}$ & 58.99 & CrNbTaTiVB$_{10}$ & 159.40 & HfMoTiVZrB$_{10}$ & 127.09 & MoTiVWZrB$_{10}$ & 111.02 \\
CrHfNbWZrB$_{10}$ & 98.65 & CrNbTaTiWB$_{10}$ & 130.74 & HfMoTiWZrB$_{10}$ & 110.41 & NbTaTiVWB$_{10}$ & 162.96 \\
CrHfTaTiVB$_{10}$ & 106.88 & CrNbTaTiZrB$_{10}$ & 91.09 & HfMoVWZrB$_{10}$ & 119.53 & NbTaTiVZrB$_{10}$ & 153.23 \\
CrHfTaTiWB$_{10}$ & 115.24 & CrNbTaVWB$_{10}$ & 165.47 & HfNbTaTiVB$_{10}$ & 193.13 & NbTaTiWZrB$_{10}$ & 154.53 \\
CrHfTaTiZrB$_{10}$ & 80.64 & CrNbTaVZrB$_{10}$ & 73.95 & HfNbTaTiWB$_{10}$ & 159.37 & NbTaVWZrB$_{10}$ & 154.87 \\
CrHfTaVWB$_{10}$ & 111.07 & CrNbTaWZrB$_{10}$ & 107.27 & HfNbTaTiZrB$_{10}$ & 240.56 & NbTiVWZrB$_{10}$ & 141.16 \\
CrHfTaVZrB$_{10}$ & 62.83 & CrNbTiVWB$_{10}$ & 158.31 & HfNbTaVWB$_{10}$ & 179.91 & TaTiVWZrB$_{10}$ & 126.13 \\
CrHfTaWZrB$_{10}$ & 94.16 & CrNbTiVZrB$_{10}$ & 81.54 & HfNbTaVZrB$_{10}$ & 137.19 & & \\
CrHfTiVWB$_{10}$ & 133.51 & CrNbTiWZrB$_{10}$ & 119.29 & HfNbTaWZrB$_{10}$ & 220.61 & & \\
\bottomrule
\end{tabular}%
}
\end{table*}

\begin{figure}[h!]
\centering
\includegraphics[width=\columnwidth]{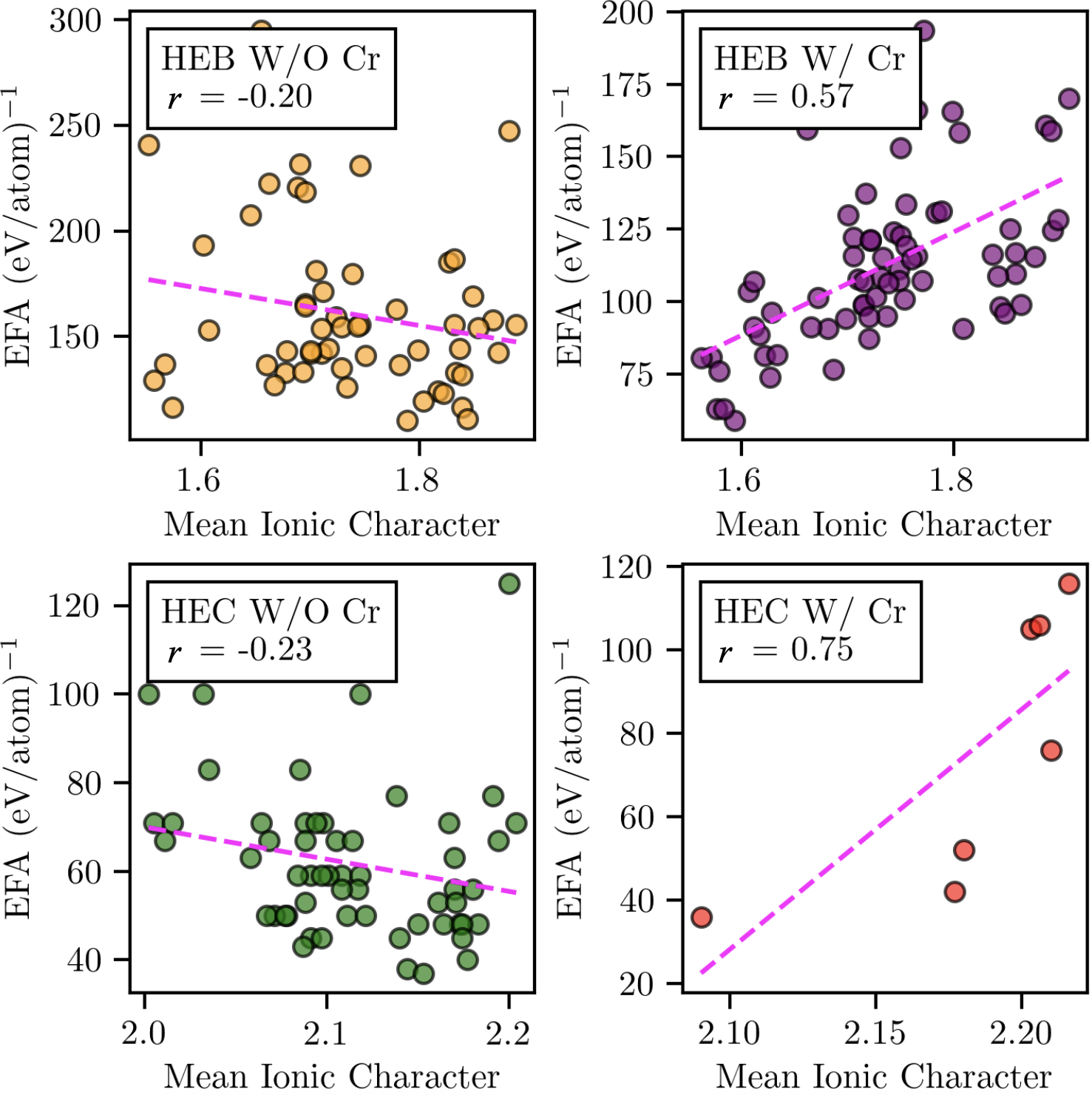}
\caption{Entropy-forming ability (EFA) versus the mean ionic character for Cr-free (left) and Cr-containing (right) high-entropy borides (HEBs, top) and high-entropy carbides (HECs, bottom). Dashed lines show linear fits, and the corresponding Pearson correlation coefficients $r$ are indicated. Cr-free and Cr-containing compositions exhibit opposite correlations between EFA and mean ionic character in both HEBs and HECs, highlighting the distinctive role of Cr in determining the EFA values. HEC data are obtained from Ref. \onlinecite{sarker2018high,kaufmann2020discovery}.
}
\label{fig:efa_ML}
\end{figure}

To further elucidate the role of Cr in the synthesizability of high-entropy ceramics, we examine the relationship between EFA and the mean ionic character $\bar{I} $ for both the five-metal hexagonal HEBs studied here and the five-metal cubic high-entropy carbides (HECs) reported previously~\cite{sarker2018high,kaufmann2020discovery}. The mean ionic character is defined as $\bar{I} = \sum_{i,j} f_i f_j \chi_i \chi_j$, where $f_i$ and $\chi_i$ represent the atomic fraction and electronegativity of element $i$ of a compound. The Pearson correlation coefficient $r$ quantifies the linear relationship between two quantities:
\begin{equation}
r =
\frac{
\sum_{i=1}^{N} \left(x_i-\bar{x}\right)\left(y_i-\bar{y}\right)
}{
\sqrt{\sum_{i=1}^{N}\left(x_i-\bar{x}\right)^2}
\sqrt{\sum_{i=1}^{N}\left(y_i-\bar{y}\right)^2}
},
\end{equation}
where $x_i$ and $y_i$ are the values of two physical quantities for sample $i$; $\bar{x}$ and $\bar{y}$ are their corresponding mean values; $N$ is the total number of samples. The Pearson correlation coefficient ranges from -1 to 1, with $r=-1$ indicating a perfect negative linear correlation, and $r=1$ a perfect positive linear correlation. Figure \ref{fig:efa_ML} shows that the EFA values exhibit a negative correlation with the mean ionic character for both HEBs and HECs that do not contain Cr. In contrast, Cr-containing compositions display the opposite trend, with EFA increasing as the mean ionic character increases. Although the number of Cr-containing HECs is limited~\cite{sarker2018high,kaufmann2020discovery}, the consistent behavior observed across both HEBs and HECs highlights the distinctive role of Cr in influencing the EFA values.

Among the transition metals considered, Cr is known for its proximity to magnetic instabilities and ordering~\cite{fawcett1988spin}. Although the present DFT calculations are performed in the non-magnetic state, local magnetic fluctuations may still influence the energy landscape of Cr-containing high-entropy ceramics. Such effects could increase the energy dispersion among competing configurations and lead to lower EFA values. Further spin-polarized calculations are needed to assess the importance of this mechanism. Nevertheless, the present results do not rule out the formation of single-phase Cr-containing HEBs at compositions away from the equimolar ratio~\cite{chakrabarty2026high}.

Finally, as discussed in the next section, several of the high-EFA compounds also exhibit favorable mechanical properties. Among the five highest-EFA HEBs, four are predicted to possess shear moduli exceeding 200 GPa and Vickers hardness values in the range of $27-35$ GPa. In particular, HfNbTaTiZrB$_{10}$ combines a high EFA value of 240.56 (eV/atom)$^{-1}$ with one of the largest calculated shear moduli (228 GPa) and hardness values (34.48 GPa), making it attractive for experimental synthesis.

\subsection{Mechanical Property}

\begin{figure*}[tbp]
\centering
\includegraphics[width=\textwidth]{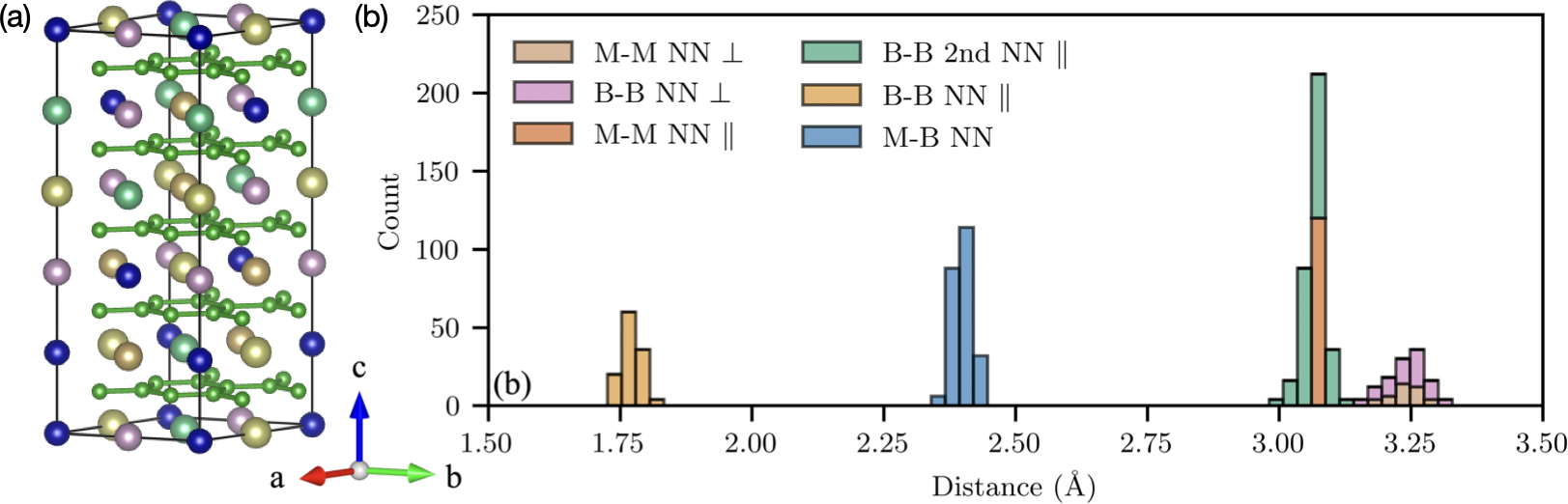}
\caption{(a) 60-atom special quasi-random structure (SQS) of hexagonal five-metal high-entropy boride (HEB), visualized by the VESTA software~\cite{momma2011vesta}. The smaller green spheres represent boron atoms, while the larger spheres are transition-metal atoms occupying the regions between hexagonal boron layers. (b) Histogram of nearest-neighbor (NN) and 2nd NN bond distances obtained from the 126 relaxed HEB SQS supercells, including boron-boron (B-B), boron-metal (B-M), and metal-metal (M-M) bonds. Bonds oriented along the $c$-axis are denoted by the symbol $\perp$, whereas bonds within the boron layers are denoted by $\parallel$.
}
\label{fig:sqs}
\end{figure*}

While EFA assesses synthesizability, mechanical properties ultimately determine the suitability of HEBs for practical applications. To capture the effects of chemical and structural disorder, we perform DFT calculations using a larger 60-atom special quasi-random structure (SQS), which is shown in Fig. \ref{fig:sqs}(a). The use of SQS supercells also enables the extraction of structural descriptors that are subsequently employed in machine-learning (ML) analysis. Figure \ref{fig:sqs}(b) presents histograms of selected structural features obtained from the 126 fully relaxed HEB SQS supercells, including nearest-neighbor (NN) and second-nearest-neighbor (2nd NN) bond distances for boron-boron (B-B), boron-metal (B-M), and metal-metal (M-M) pairs. The bonds are oriented either along the crystal $c$-axis (denoted by $\perp$) or within the hexagonal boron layers (denoted by $\parallel$). The bond-length distributions are approximately Gaussian, suggesting that their mean values and standard deviations can serve as structural descriptors for ML models.

Table \ref{table:mech} summarizes the DFT-calculated mechanical properties of all 126 five-metal HEBs studied. The reported quantities include the bulk modulus ($B_{DFT}$), shear modulus ($G_{DFT}$), and Vickers hardness ($H_{DFT}$) estimated from Tian's empirical model~\cite{tian2012microscopic}. Materials marked with an asterisk (*) in the Table indicate a large discrepancy between the Voigt and Reuss elastic moduli (over 50 GPa), while compounds marked with ``-'' did not achieve satisfactory convergence in the elastic-constant calculations. The Voigt and Reuss approximations respectively provide upper and lower bounds for the elastic moduli~\cite{hill1952elastic,ravindran1998density}. A small Voigt-Reuss difference generally indicates nearly isotropic and mechanically stable behavior. A large discrepancy suggests significant elastic anisotropy or proximity to mechanical instability, and these entries are excluded from the ML dataset. It is also noted that out of the 10 compounds labeled with * or ``-'', 80\% of them contain Cr. The mechanical instability is likely associated with the Cr-induced severe lattice distortion, as discussed in a recent DFT-SQS study~\cite{mo2026chromium}. These results reveal that Cr-containing HEBs exhibit more complex behavior and are more prone to mechanical instability or convergence difficulties in DFT calculations.

\begin{table*}[t]
\caption{Calculated bulk modulus ($B_{DFT}$), shear modulus ($G_{DFT}$), and Vickers hardness ($H_{DFT}$) in units of GPa for the 126 five-metal high-entropy borides studied. Materials marked with an asterisk (*) were excluded from the machine-learning dataset due to a Voigt-Reuss difference greater than 50 GPa in $B$ and/or $G$. Entries marked with ``-'' did not converge.}
\label{table:mech}
\centering
\small
\setlength{\tabcolsep}{3.25pt}
\setlength{\arrayrulewidth}{0.8pt}
\resizebox{\textwidth}{!}{%
\begin{tabular}{lrrr|lrrr|lrrr}
\toprule
Material & B$_{DFT}$ & G$_{DFT}$ & H$_{DFT}$ & Material & B$_{DFT}$ & G$_{DFT}$ & H$_{DFT}$ & Material & B$_{DFT}$ & G$_{DFT}$ & H$_{DFT}$ \\
\midrule
CrHfMoNbTaB$_{10}$ &    291.47 &    179.77 &     20.97 & *CrMoNbVWB$_{10}$   &    308.04 &    217.05 &     27.87 & HfMoTaVZrB$_{10}$  &    281.94 &    206.98 &     28.24 \\
CrHfMoNbTiB$_{10}$ &    283.38 &    195.02 &     25.15 & CrMoNbVZrB$_{10}$  &    282.70 &    153.73 &     16.26 & HfMoTaWZrB$_{10}$  &    293.25 &    190.64 &     23.20 \\
CrHfMoNbVB$_{10}$  &    288.09 &    179.66 &     21.22 & *CrMoNbWZrB$_{10}$  &    222.38 &    147.11 &     19.70 & HfMoTiVWB$_{10}$   &    296.25 &    203.98 &     25.98 \\
CrHfMoNbWB$_{10}$  &    288.97 &    144.56 &     14.16 & CrMoTaTiVB$_{10}$  &    300.28 &    190.10 &     22.47 & HfMoTiVZrB$_{10}$  &    272.15 &    214.37 &     31.36 \\
CrHfMoNbZrB$_{10}$ &    274.63 &    190.11 &     24.87 & CrMoTaTiWB$_{10}$  &    303.85 &    147.70 &     13.92 & HfMoTiWZrB$_{10}$  &    283.67 &    202.10 &     26.83 \\
CrHfMoTaTiB$_{10}$ &    288.76 &    197.18 &     25.13 & CrMoTaTiZrB$_{10}$ &    285.12 &    182.32 &     22.06 & HfMoVWZrB$_{10}$   &    286.70 &    194.03 &     24.59 \\
CrHfMoTaVB$_{10}$  &    293.49 &    179.66 &     20.78 & *CrMoTaVWB$_{10}$   &    300.49 &    113.98 &      8.74 & HfNbTaTiVB$_{10}$  &    286.36 &    226.58 &     32.78 \\
CrHfMoTaWB$_{10}$  &    290.91 &    140.84 &     13.39 & *CrMoTaVZrB$_{10}$  &    288.30 &    247.16 &     38.19 & HfNbTaTiWB$_{10}$  &    297.43 &    209.77 &     27.24 \\
CrHfMoTaZrB$_{10}$ &    279.96 &    192.33 &     24.86 & *CrMoTaWZrB$_{10}$  &    276.73 &    104.40 &      8.16 & HfNbTaTiZrB$_{10}$ &    276.71 &    228.01 &     34.48 \\
CrHfMoTiVB$_{10}$  &    285.46 &    201.10 &     26.40 & CrMoTiVWB$_{10}$   &    306.76 &    172.39 &     18.31 & HfNbTaVWB$_{10}$   &    301.44 &    202.69 &     25.18 \\
CrHfMoTiWB$_{10}$  &    294.21 &    155.83 &     15.94 & CrMoTiVZrB$_{10}$  &    278.98 &    203.93 &     27.81 & HfNbTaVZrB$_{10}$  &    279.29 &    218.74 &     31.61 \\
CrHfMoTiZrB$_{10}$ &    271.34 &    206.05 &     29.25 & CrMoTiWZrB$_{10}$  &    -      &    -      &     -     & HfNbTaWZrB$_{10}$  &    290.48 &    202.14 &     26.13 \\
CrHfMoVWB$_{10}$   &    299.06 &    179.99 &     20.41 & CrMoVWZrB$_{10}$   &    290.83 &    175.71 &     20.15 & HfNbTiVWB$_{10}$   &    291.59 &    216.21 &     29.46 \\
CrHfMoVZrB$_{10}$  &    274.62 &    188.50 &     24.49 & CrNbTaTiVB$_{10}$  &    293.22 &    213.73 &     28.65 & HfNbTiVZrB$_{10}$  &    267.61 &    225.30 &     35.04 \\
CrHfMoWZrB$_{10}$  &    281.84 &    181.30 &     22.12 & *CrNbTaTiWB$_{10}$  &    300.24 &    329.33 &     61.94 & HfNbTiWZrB$_{10}$  &    280.27 &    214.44 &     30.35 \\
CrHfNbTaTiB$_{10}$ &    285.24 &    212.09 &     29.15 & CrNbTaTiZrB$_{10}$ &    279.25 &    212.61 &     30.00 & HfNbVWZrB$_{10}$   &    283.04 &    206.31 &     27.95 \\
CrHfNbTaVB$_{10}$  &    288.83 &    197.74 &     25.25 & CrNbTaVWB$_{10}$   &    307.53 &    181.70 &     20.12 & HfTaTiVWB$_{10}$   &    296.96 &    218.49 &     29.42 \\
CrHfNbTaWB$_{10}$  &    299.40 &    191.09 &     22.76 & CrNbTaVZrB$_{10}$  &    282.64 &    198.46 &     26.06 & HfTaTiVZrB$_{10}$  &    273.05 &    228.62 &     35.18 \\
CrHfNbTaZrB$_{10}$ &    276.53 &    206.40 &     28.72 & CrNbTaWZrB$_{10}$  &    273.74 &    139.47 &     14.10 & HfTaTiWZrB$_{10}$  &    285.08 &    216.34 &     30.26 \\
CrHfNbTiVB$_{10}$  &    278.67 &    214.77 &     30.63 & CrNbTiVWB$_{10}$   &    -      &    -      &     -     & HfTaVWZrB$_{10}$   &    287.96 &    208.33 &     27.90 \\
CrHfNbTiWB$_{10}$  &    290.20 &    194.36 &     24.33 & CrNbTiVZrB$_{10}$  &    272.52 &    210.68 &     30.33 & HfTiVWZrB$_{10}$   &    278.06 &    218.14 &     31.61 \\
CrHfNbTiZrB$_{10}$ &    267.32 &    214.48 &     32.04 & CrNbTiWZrB$_{10}$  &    283.56 &    200.54 &     26.47 & MoNbTaTiVB$_{10}$  &    299.27 &    208.56 &     26.76 \\
CrHfNbVWB$_{10}$   &    294.39 &    183.01 &     21.42 & CrNbVWZrB$_{10}$   &    286.91 &    188.56 &     23.31 & MoNbTaTiWB$_{10}$  &    308.17 &    188.27 &     21.43 \\
CrHfNbVZrB$_{10}$  &    269.84 &    204.45 &     29.02 & CrTaTiVWB$_{10}$   &    304.96 &    205.42 &     25.47 & MoNbTaTiZrB$_{10}$ &    286.93 &    209.60 &     28.33 \\
CrHfNbWZrB$_{10}$  &    280.76 &    194.90 &     25.40 & CrTaTiVZrB$_{10}$  &    277.95 &    213.27 &     30.33 & *MoNbTaVWB$_{10}$   &    311.01 &    122.95 &      9.66 \\
CrHfTaTiVB$_{10}$  &    284.26 &    218.12 &     30.82 & CrTaTiWZrB$_{10}$  &    288.48 &    202.29 &     26.37 & MoNbTaVZrB$_{10}$  &    289.61 &    200.79 &     25.90 \\
CrHfTaTiWB$_{10}$  &    295.30 &    196.11 &     24.25 & CrTaVWZrB$_{10}$   &    292.18 &    190.07 &     23.17 & MoNbTaWZrB$_{10}$  &    299.69 &    186.08 &     21.65 \\
CrHfTaTiZrB$_{10}$ &    272.53 &    217.67 &     32.21 & CrTiVWZrB$_{10}$   &    283.25 &    187.72 &     23.46 & MoNbTiVWB$_{10}$   &    305.43 &    196.97 &     23.53 \\
CrHfTaVWB$_{10}$   &    299.52 &    188.03 &     22.08 & HfMoNbTaTiB$_{10}$ &    292.89 &    211.78 &     28.21 & MoNbTiVZrB$_{10}$  &    280.08 &    209.26 &     29.03 \\
CrHfTaVZrB$_{10}$  &    275.13 &    207.60 &     29.19 & HfMoNbTaVB$_{10}$  &    295.31 &    202.57 &     25.75 & MoNbTiWZrB$_{10}$  &    290.51 &    195.53 &     24.57 \\
CrHfTaWZrB$_{10}$  &    285.65 &    196.41 &     25.26 & HfMoNbTaWB$_{10}$  &    304.21 &    188.33 &     21.76 & MoNbVWZrB$_{10}$   &    293.05 &    183.63 &     21.67 \\
CrHfTiVWB$_{10}$   &    290.79 &    199.38 &     25.45 & HfMoNbTaZrB$_{10}$ &    285.65 &    205.60 &     27.48 & MoTaTiVWB$_{10}$   &    310.71 &    197.64 &     23.22 \\
CrHfTiVZrB$_{10}$  &    264.70 &    214.85 &     32.50 & HfMoNbTiVB$_{10}$  &    285.96 &    211.41 &     28.89 & MoTaTiVZrB$_{10}$  &    285.36 &    211.86 &     29.08 \\
CrHfTiWZrB$_{10}$  &    276.81 &    207.95 &     29.08 & HfMoNbTiWB$_{10}$  &    296.32 &    183.11 &     21.29 & MoTaTiWZrB$_{10}$  &    295.51 &    197.60 &     24.57 \\
CrHfVWZrB$_{10}$   &    279.39 &    195.01 &     25.56 & HfMoNbTiZrB$_{10}$ &    275.33 &    213.35 &     30.68 & MoTaVWZrB$_{10}$   &    297.89 &    185.54 &     21.68 \\
CrMoNbTaTiB$_{10}$ &    299.15 &    192.66 &     23.13 & HfMoNbVWB$_{10}$   &    298.47 &    187.04 &     21.96 & MoTiVWZrB$_{10}$   &    290.06 &    199.82 &     25.62 \\
CrMoNbTaVB$_{10}$  &    303.51 &    160.28 &     16.20 & HfMoNbVZrB$_{10}$  &    277.18 &    204.96 &     28.27 & NbTaTiVWB$_{10}$   &    304.64 &    211.10 &     26.82 \\
CrMoNbTaWB$_{10}$  &    301.14 &    124.83 &     10.31 & HfMoNbWZrB$_{10}$  &    288.68 &    189.93 &     23.46 & NbTaTiVZrB$_{10}$  &    280.29 &    223.19 &     32.67 \\
CrMoNbTaZrB$_{10}$ &    289.49 &    186.60 &     22.63 & HfMoTaTiVB$_{10}$  &    291.15 &    214.07 &     28.97 & NbTaTiWZrB$_{10}$  &    291.40 &    205.75 &     26.90 \\
CrMoNbTiVB$_{10}$  &    294.66 &    189.47 &     22.82 & *HfMoTaTiWB$_{10}$  &    301.06 &     92.44 &      5.92 & NbTaVWZrB$_{10}$   &    293.91 &    198.60 &     24.96 \\
CrMoNbTiWB$_{10}$  &    301.26 &    170.05 &     18.22 & HfMoTaTiZrB$_{10}$ &    280.31 &    215.92 &     30.73 & NbTiVWZrB$_{10}$   &    285.14 &    211.83 &     29.09 \\
CrMoNbTiZrB$_{10}$ &    279.70 &    189.59 &     24.24 & HfMoTaVWB$_{10}$   &    303.46 &    185.83 &     21.29 & TaTiVWZrB$_{10}$   &    290.15 &    213.67 &     28.98 \\
\bottomrule
\end{tabular}%
}
\end{table*}

\begin{figure}[tbp]
\centering
\includegraphics[width=\columnwidth]{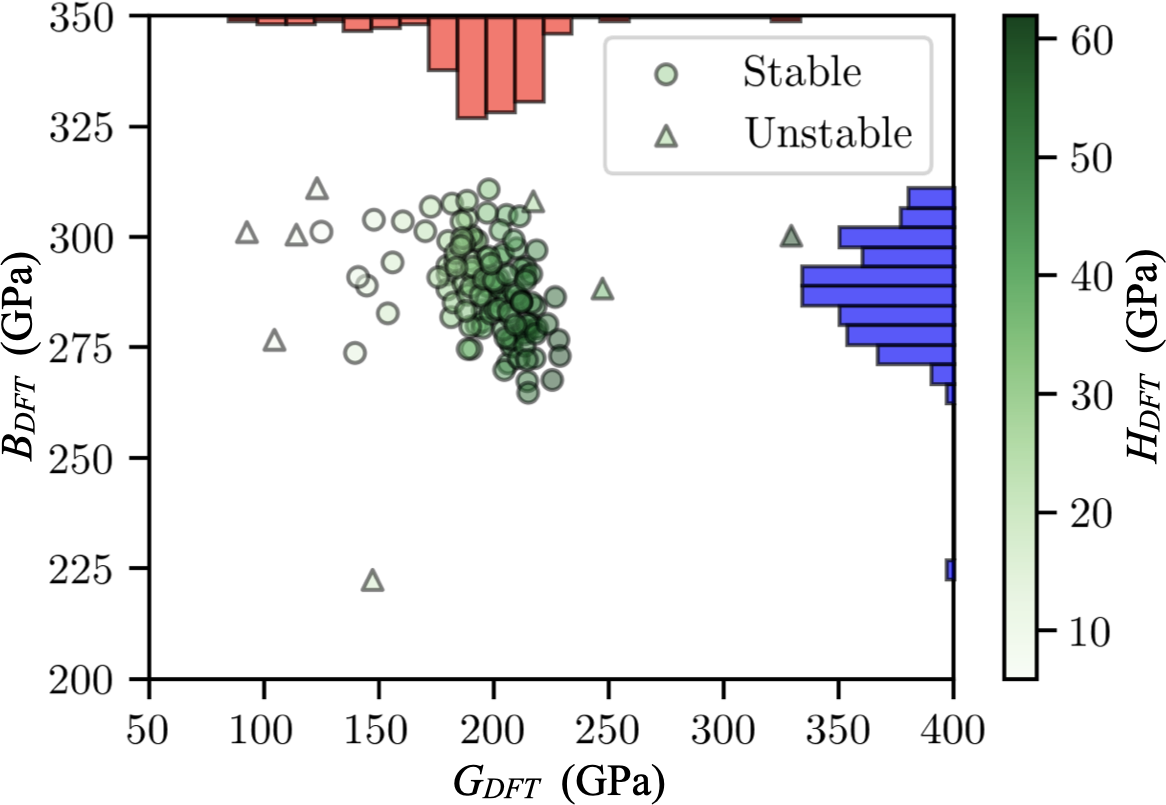}
\caption{Joint scatter plot and marginal histograms of the DFT-calculated shear modulus ($G_{DFT}$) and bulk modulus ($B_{DFT}$) for the high-entropy borides studied. The top and right histograms show the distributions of ($G_{DFT}$, red) and ($B_{DFT}$, blue), respectively. The color scale represents the Vickers hardness ($H_{DFT}$, green) estimated using Tian's empirical model~\cite{tian2012microscopic}. Circles denote mechanically stable compounds, while triangles denote mechanically unstable ones.
}
\label{fig:mech_DFT}
\end{figure}

To better visualize the DFT-calculated mechanical properties, Fig. \ref{fig:mech_DFT} presents the distributions and correlations of $B_{DFT}$, $G_{DFT}$, and $H_{DFT}$. The majority of mechanically stable compounds cluster within a relatively narrow range of bulk moduli between approximately $270-310$ GPa, indicating that the resistance to volume compression is fairly insensitive to composition. In contrast, the shear modulus exhibits a much broader distribution, spanning roughly $140-230$ GPa for most stable compounds, suggesting a stronger dependence on chemical composition and local bonding characteristics. A small number of mechanically unstable compounds, denoted by triangles, fall outside the main cluster and are characterized by unusually low or high elastic moduli. The relatively high bulk and shear moduli of the HEBs reflect the intrinsically stiff bonding network of transition-metal borides. Interestingly, $B_{DFT}$ and $G_{DFT}$ exhibit a weak negative correlation, with compounds possessing larger shear moduli tending to have somewhat lower bulk moduli. The calculated hardness values also vary substantially, ranging from about $10-60$ GPa, although the majority of mechanically stable compounds fall between roughly $15- 35$ GPa. Since empirical hardness models depend more strongly on the shear modulus than on the bulk modulus, compounds with larger $G_{DFT}$ generally exhibit higher hardness values~\cite{teter1998computational,chen2011modeling}.

Finally, we employ machine-learning (ML) models to analyze the first-principles mechanical-property data. Three separate XGBoost models are trained to predict $B_{DFT}$, $G_{DFT}$, and $H_{DFT}$, excluding the 10 compounds marked with * or ``--'' in Table \ref{table:mech}. To evaluate model performance, we use both the Pearson correlation coefficient $r$ between the ML-predicted values ($\hat{y}_i$) and the corresponding DFT-calculated values ($y_i$), as well as the mean absolute error (MAE), defined as $\text{MAE} =  (\sum^N_{i=1} |y_i - \hat{y_i}|)/N$, where $N$ is the number of samples. While $r$ quantifies the linear correlation strength between the ML predictions and DFT results, MAE measures the average prediction error in units of GPa, providing a more direct assessment of the ML model performance.

\begin{figure*}[tbp]
\centering
\includegraphics[width=0.8\textwidth]{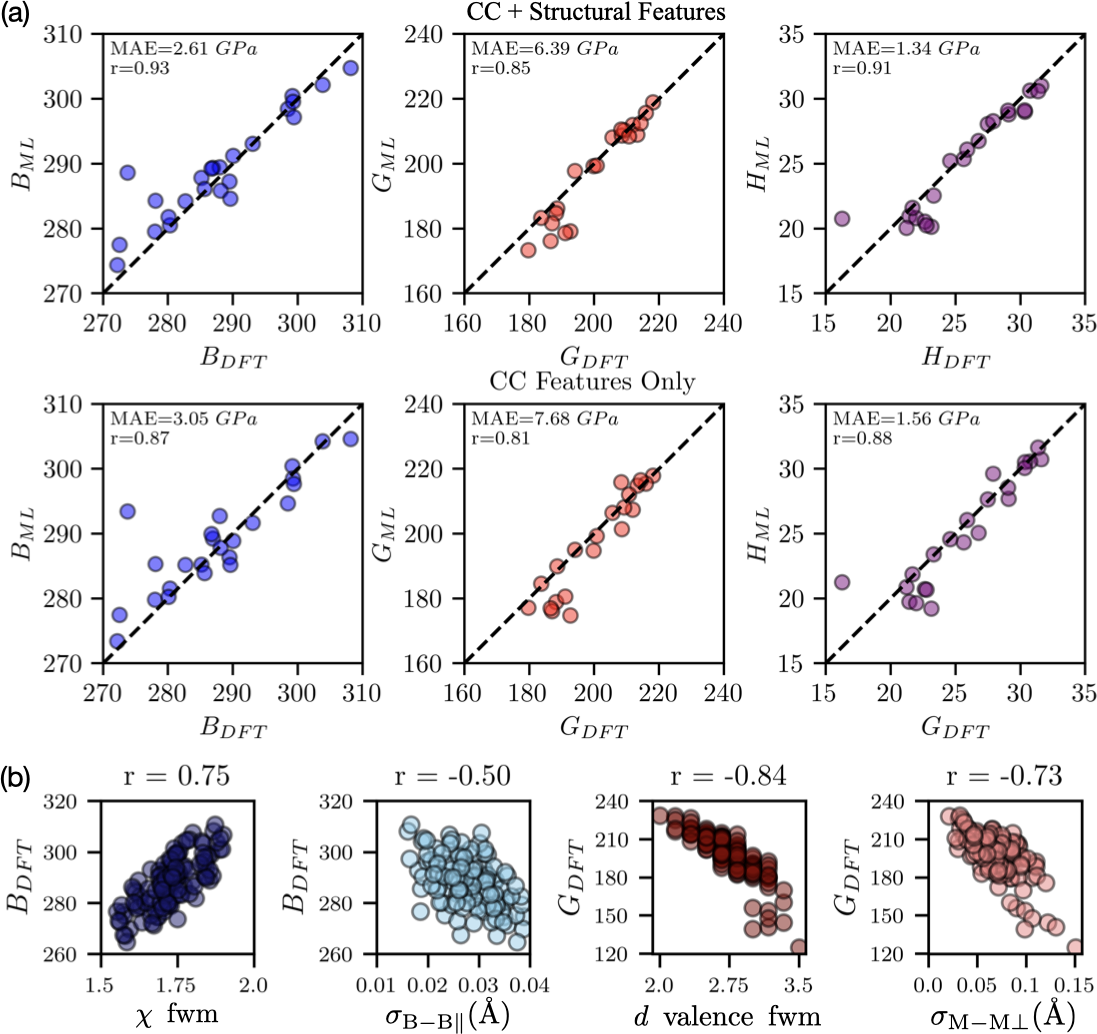}
\caption{(a) DFT-calculated bulk modulus ($B_{DFT}$), shear modulus ($G_{DFT}$), and Vickers hardness ($H_{DFT}$), compared with the machine learning (ML) model predictions. Models in the top and bottom rows use composition-based (CC) plus structural descriptors and CC descriptors only, respectively. The Pearson correlation coefficient ($r$) and mean absolute error (MAE) are indicated in each panel. The dashed line denotes perfect agreement. (b) Correlations between the mechanical properties and the most important CC and structural descriptors identified by XGBoost. The corresponding $r$ values are shown for each plot. ``$\chi$ fwm" stands for the fractional weighted mean of electronegativity. $\sigma$ denotes the standard deviation.
}
\label{fig:mech_ML}
\end{figure*}

Figure \ref{fig:mech_ML}(a) compares the ML-predicted and DFT-calculated bulk moduli, shear moduli, and Vickers hardness values. When both composition-based (CC) and structural descriptors are included, the XGBoost models achieve strong predictive performance, with Pearson correlation coefficients of $r=0.93$, $0.85$, and $0.91$, and MAE values of 2.61, 6.39, and 1.34 GPa for $B_{DFT}$, $G_{DFT}$, and $H_{DFT}$, respectively. Models trained using only CC descriptors exhibit slightly lower accuracy, with $r=0.87$, 0.81, and 0.88, and MAE values of 3.05, 7.68, and 1.56 GPa. These results indicate that composition-based descriptors alone capture much of the variation in the mechanical properties, while the inclusion of structural descriptors extracted from the SQS supercells provides a noticeable improvement in the prediction accuracy.

Figure \ref{fig:mech_ML}(b) shows the correlations between the mechanical properties and the most important CC and structural features identified by the XGBoost models. Among the CC descriptors, the fractional weighted mean (fwm) of the elemental electronegativity ($\chi$) exhibits a strong positive correlation with the bulk modulus ($r=0.75$), while the fwm of the $d$-valence occupation shows a strong negative correlation with the shear modulus ($r=-0.84$). The positive correlation between the variation in $\chi$ and bulk modulus suggests that greater chemical contrast among the constituent elements may enhance the bond stiffness and resistance to compression. In contrast, variations in $d$-electron filling alter the occupancy of bonding and antibonding states, thereby affecting the resistance to shear deformation. Larger variations in $d$-electron occupation tend to reduce the shear modulus.

Among the structural descriptors, the standard deviation of the in-plane B-B bond lengths ($\sigma_{\mathrm{B-B}\parallel}$) is negatively correlated with the bulk modulus ($r=-0.50$), and the standard deviation of the out-of-plane metal-metal bond lengths ($\sigma_{\mathrm{M-M}\perp}$) is negatively correlated with the shear modulus ($r=-0.73$). These trends suggest that both compositional diversity and structural disorder play important roles. The ML results further suggest that optimizing bulk and shear moduli in HEBs may require different design strategies, as they are governed by distinct compositional and structural features.

Finally, we note that the relatively small dataset of 126 samples limits the development of highly accurate and broadly transferable ML models. While the models achieve reasonably good performance, our primary objective is not to maximize the prediction accuracy, but rather to identify the key compositional and structural features governing the mechanical properties of HEBs, and to assess whether descriptors related to lattice disorder contribute meaningfully to the ML analysis.
 
\section{Conclusion}\label{conclusion}
We have performed a systematic density functional theory (DFT) and machine-learning (ML) study of 126 five-metal high-entropy borides (HEBs) in the hexagonal AlB$_2$ structure. The entropy-forming ability (EFA) and elastic properties were evaluated from first principles, while XGBoost models were developed to identify the key compositional and structural descriptors governing mechanical performance. The EFA analysis revealed substantial composition-dependent variations and demonstrated good agreement with the x-ray diffraction measurements obtained in this study, confirming the usefulness of EFA for assessing single-phase synthesizability. A notable finding is that Cr-containing compositions generally exhibiting lower EFA values and account for the majority of compounds displaying mechanical instability or convergence difficulties in DFT calculations. The calculated mechanical properties show that the HEBs possess consistently high bulk and shear moduli characteristic of transition-metal borides. The ML analysis also revealed  important compositional features and structural descriptors associated with local bond-length variations, providing further improvements in the ML prediction.

Overall, the combined DFT and ML framework provides a computational roadmap for the design and synthesis of single-phase HEBs with superior mechanical performances. Further experimental validation of the predicted EFA trends and mechanical properties across a broader range of compositions can be important future work. It will also be valuable to extend the present methodology to other high-entropy ceramic families, including carbides, silicides, and oxides, as well as to compositionally more complex systems containing six or more principal elements. Such studies will further advance our understanding of the interplay between configurational entropy, local structural disorder, synthesizability, and functional properties in high-entropy materials.

\section*{Data and Code Availability Statement}
CSV files containing the mechanical property data and EFA results are publicly available in the \texttt{properties} directory of the project's \href{https://github.com/ethanfox2206/HEB_ML}{GitHub repository}. Fully relaxed SQS supercells are provided in the \texttt{CONTCARS} directory. The machine-learning models are included within the same repository.

\section*{acknowledgments}
This work is supported by the National Science Foundation (NSF) EPSCoR Program under Award No. OIA-2148653 (FTPP).
L.M. acknowledges support from the NASA-Alabama Space Grant Consortium Training Grant 90NSSC20M0044. E.F. and J.R.P. acknowledge support from the NSF REU site at the University of Alabama at Birmingham under Award No. DMR-2445516.
B.S. and S.A.C. acknowledge support from NSF Awards No. DMR-2203112 and DMR-2116564.
The calculations are performed on the Texas Advanced Computing Center (TACC) Frontera system, which is made possible by NSF Award No. OAC-1818253.


%

\end{document}